\documentclass[a4paper,11pt]{article}
\usepackage{pos}

\usepackage{lineno}

\title{Investigating baryon-strangeness and charge-strangeness correlations in Pb--Pb collisions at $\sqrt{s_\mathrm{NN}}$ = 5.02 TeV with ALICE}
\ShortTitle{Baryon-strangeness and charge-strangeness correlations}

\author*[a]{Swati Saha (for the ALICE Collaboration)}

\affiliation[a]{National Institute of Science Education and Research, Jatni 752050, India\\
  Homi Bhabha National Institute, Training School Complex, Anushaktinagar, Mumbai 400094, India}

\emailAdd{swati.saha@cern.ch}
\emailAdd{swati.saha@niser.ac.in}

\abstract{
  To explore the quantum chromodynamics (QCD) phase transitions and the properties of quark--gluon plasma, the ALICE collaboration at CERN has conducted an extensive analysis of the correlations among net-conserved quantities, namely net-baryon, net-charge, and net-strangeness. These correlations are essential for understanding the QCD phase structure, as they are directly connected to ratios of thermodynamic susceptibilities calculated in lattice QCD. This analysis focuses on the correlations between net-kaon and net-proton, as well as net-kaon and net-charge, in Pb--Pb collisions at $\sqrt{s_\mathrm{NN}} = 5.02$ TeV, where net-proton and net-kaon serve as effective proxies for net-baryon and net-strangeness, respectively. A comparison with theoretical predictions from the Thermal-FIST model sheds light on the role of resonance decays and the effects of charge conservation laws in shaping these correlations. Furthermore, the measurements show sensitivity to the correlation volume in which these conservation laws are applied, underscoring the importance of modeling the underlying dynamics to fully understand the experimental results on fluctuations and correlations in heavy-ion collisions.
  } 

\FullConference{%
  42nd International Conference on High Energy physics - ICHEP2024\\
  18-24 July, 2024\\
  Prague, Czech Republic
}

\begin{document}
\maketitle


\section{Introduction}
\label{intro}
The quark--gluon plasma (QGP), a deconfined state of matter predicted by Quantum Chromodynamics (QCD), emerges at high energy densities and temperature in heavy-ion collisions \cite{Intro1}. A key tool for probing the QGP and the QCD phase transition is the analysis of fluctuations and correlations of conserved charges: baryon number (B), electric charge (Q), and strangeness (S) \cite{Motiv1, Motiv2}. These observables are directly related to thermodynamic susceptibilities as calculated within lattice QCD (LQCD), and offer insight into the QCD phase structure, particularly the crossover between the hadronic phase and the QGP phase \cite{Friman, FKarsch}. Comparison of measurements on fluctuations and correlations of conserved charges with Hadron Resonance Gas (HRG) model predictions can also provide crucial constraints on the freeze-out conditions and the thermodynamic properties of the system created in heavy-ion collisions \cite{FKarsch}. However, observed fluctuations are also influenced by several other factors such as quantum number conservation laws  \cite{Qua}, resonance decays \cite{Reso2}, volume fluctuations \cite{Vol2} etc., all of which must be disentangled to extract meaningful QCD-related signals. This study investigates how the conservation of B, Q, and S impacts particle production, influencing both fluctuations and correlations, while also examining the role of resonance decays and post-hadronization interactions. Furthermore, the analysis reveals a sensitivity of these fluctuations and correlations to the finite volume where conservation laws are applied, highlighting the importance of accurately modeling spatial constraints. Through detailed experimental measurements, compared with predictions from the Thermal-FIST model, these effects are quantified to refine the understanding of how conservation laws and underlying dynamics shape the behavior of the QCD medium in heavy-ion collisions.

In LQCD, the thermodynamic susceptibilities of conserved charges (B, Q, S) in a system with volume $V$ and temperature $T$, are derived from the partial derivatives of the reduced QCD pressure, $P/T^{4}$, with respect to the chemical potentials associated with baryon number ($\mu_\mathrm{B}$), electric charge ($\mu_\mathrm{Q}$), and strangeness ($\mu_\mathrm{S}$)
\begin{equation}
  \chi^{l,m,n}_\mathrm{B,Q,S} = \frac{\partial^{(l+m+n)}(P/T^{4})}{\partial^{l}(\mu_\mathrm{B}/T)\partial^{m}(\mu_\mathrm{Q}/T)\partial^{n}(\mu_\mathrm{S}/T)},
\end{equation}
where $l, m, n$ represents the order of derivative. These susceptibilities are directly connected to the cumulants of the net-multiplicity distributions for B, Q, and S. The cumulants, denoted by $\sigma^{l,m,n}_\mathrm{B,Q,S}$, are related to the susceptibilities as $\chi^{l,m,n}_\mathrm{B,Q,S} = \frac{1}{VT^3}\sigma^{l,m,n}_\mathrm{B,Q,S}$. Due to experimental difficulties in detecting all baryons and strange hadrons, net-proton (p) and net-kaon (K) are used as proxies for net-baryon and net-strangeness, respectively. To eliminate the volume and temperature dependence, susceptibilities are analyzed through ratios of cumulants of net-conserved charge distributions. Specifically, the second-order ($l+m+n=2$) cumulant ratios examined here are
\begin{equation}
  C_\mathrm{p,K}=\frac{\sigma^{11}_\mathrm{p,K}}{\sigma^{2}_\mathrm{K}}=\frac{\langle(N_{p}-N_{\bar{p}})(N_{K^{+}}-N_{K^{-}})\rangle - \langle(N_{p}-N_{\bar{p}})\rangle\langle(N_{K^{+}}-N_{K^{-}})\rangle}{\langle(N_{K^{+}}-N_{K^{-}})^2\rangle - \langle(N_{K^{+}}-N_{K^{-}})\rangle^2},
\end{equation}
\begin{equation}
  C_\mathrm{Q,K}=\frac{\sigma^{11}_\mathrm{Q,K}}{\sigma^{2}_\mathrm{K}}=\frac{\langle(N_{Q^+}-N_{Q^-})(N_{K^{+}}-N_{K^{-}})\rangle - \langle(N_{Q^+}-N_{Q^-})\rangle\langle(N_{K^{+}}-N_{K^{-}})\rangle}{\langle(N_{K^{+}}-N_{K^{-}})^2\rangle - \langle(N_{K^{+}}-N_{K^{-}})\rangle^2},
\end{equation}
where $N_{Q^+}$ ($N_{Q^-}$) equals to $N_{\pi^+}+N_{K^+}+N_{p}$ ($N_{\pi^-}+N_{K^-}+N_{\bar{p}}$) and angular brackets $\langle...\rangle$ represent average over all events. This report presents measurements of $C_\mathrm{p,K}$ and $C_\mathrm{Q,K}$ as functions of centrality in Pb--Pb collisions at $\sqrt{s_\mathrm{NN}} = 5.02$ TeV, utilizing data from the ALICE detector \cite{Alice}.

\section{Analysis details}
\label{expt}
\begin{figure}[h]
\centering
\includegraphics[width=0.46\linewidth]{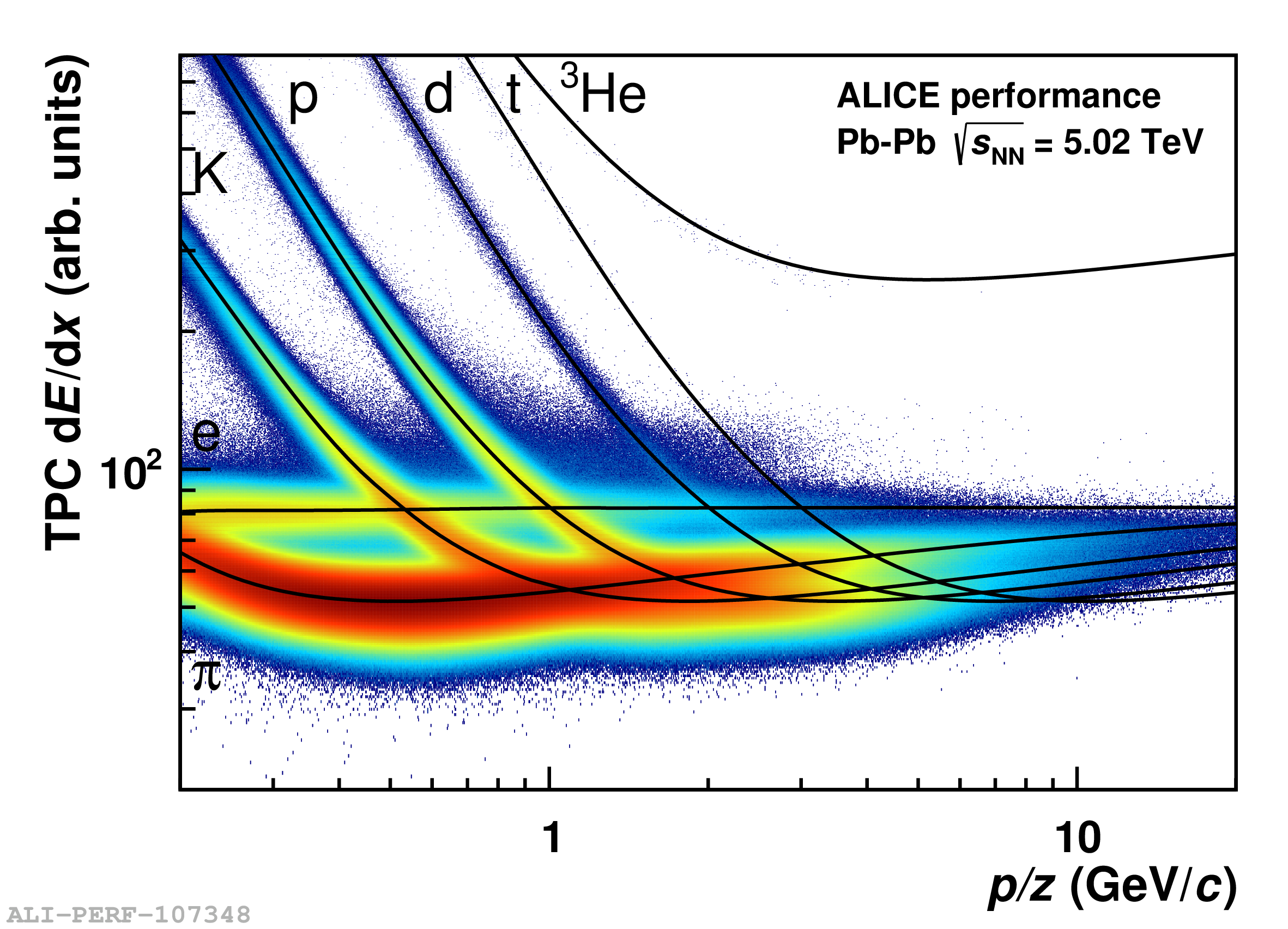}
\includegraphics[width=0.46\linewidth]{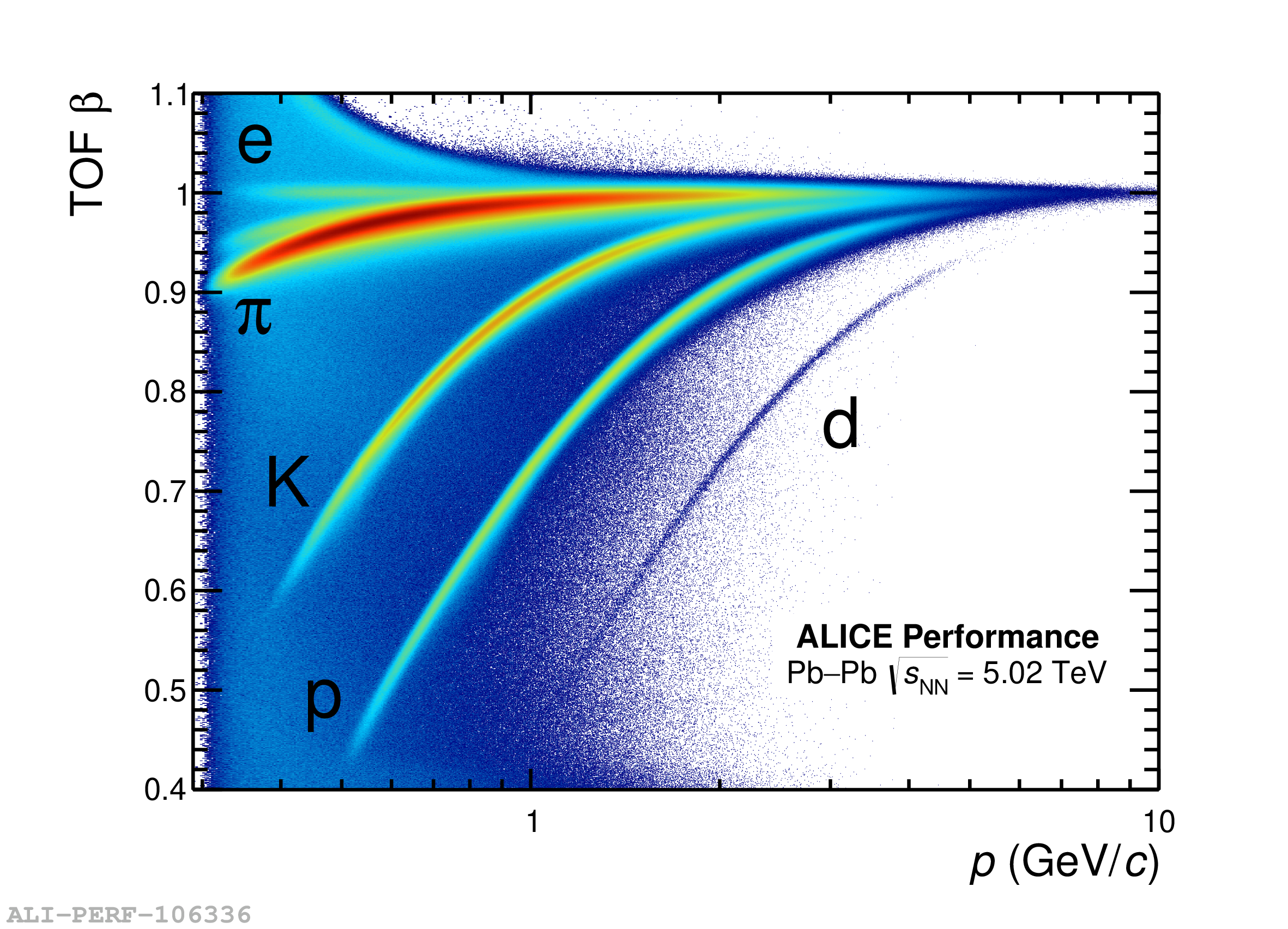}
\caption{TPC $\mathrm{d}E/\mathrm{d}x$ spectrum vs. momentum divided by the charge number, $p/z$ with superimposed Bethe--Bloch lines (left) and TOF $\beta$ vs. momentum, $p$ (right) for various particle species in Pb--Pb collisions at $\sqrt{s_\mathrm{NN}} = 5.02$ TeV.}
\label{fig-PID}       
\end{figure}

The analysis utilizes a dataset of 80 million Pb--Pb collision events selected using a minimum-bias trigger that requires signals from both the V0A detector (covering the pseudorapidity range $2.8 < \eta < 5.1$) and the V0C detector (covering $-3.7 < \eta < -1.7$). Event centrality is determined based on the signal amplitudes from these detectors. Charged particle tracks are reconstructed using the Inner Tracking System (ITS) and the Time Projection Chamber (TPC), which offer full azimuthal coverage within the pseudorapidity range $|\eta| < 0.8$. Particle identification for pions, kaons, and protons is performed by analyzing specific energy loss ($\mathrm{d}E/\mathrm{d}x$) in the TPC (see left panel of Fig.~\ref{fig-PID}) and by measuring the time of flight from the primary vertex to the Time-of-Flight (TOF) detector (see right panel of Fig.~\ref{fig-PID}). The transverse momentum ($p_\mathrm{T}$) ranges for particle selection are 0.2 $< p_\mathrm{T} <$ 2.0 GeV/$c$ for pions and kaons, and 0.4 $< p_\mathrm{T} <$ 2.0 GeV/$c$ for protons. The measurements are corrected for detector inefficiencies using the analytic relations provided in Ref.~\cite{Eff}. Track reconstruction efficiencies for protons, pions, and kaons (including their antiparticles) are obtained from Monte Carlo simulations using HIJING \cite{hijing} and ALICE detector configuration implemented through GEANT3 \cite{Geant}, and these efficiencies are applied to correct the observables. Statistical uncertainties are estimated using the bootstrap resampling technique, while systematic uncertainties are assessed by varying event and track selection criteria and by adjusting particle identification conditions.

\section{Results}
\label{results}
\begin{figure}
\centering
\includegraphics[width=0.84\linewidth]{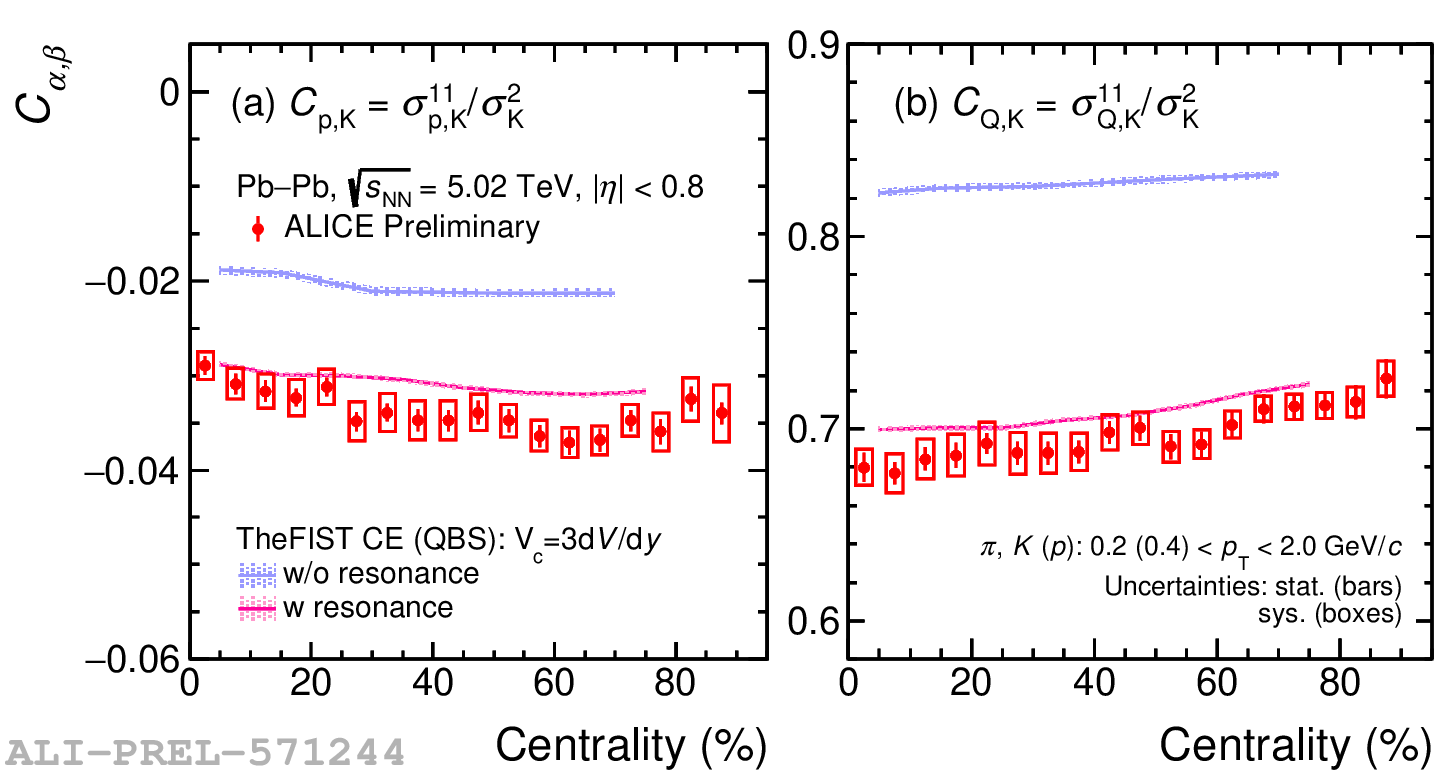}
\caption{Centrality dependence of $C_\mathrm{p,K}$ (a) and $C_\mathrm{Q,K}$ (b) in Pb--Pb collisions at $\sqrt{s_\mathrm{NN}} = 5.02$ TeV, compared with Thermal-FIST (TheFIST) \cite{TheFIST} predictions using the canonical ensemble (CE) formalism, enforcing exact conservation of baryon number, electric charge, and strangeness (B, Q, S) within a correlation volume of $V_{c} = 3\mathrm{d}V/\mathrm{d}y$. ALICE data are shown as solid red circles, while colored bands represent Thermal-FIST calculations with (w) and without (w/o) resonance contributions.}
\label{fig-1}       
\end{figure}

The measured correlation between net-proton and net-kaon, denoted as $C_\mathrm{p,K}$, is shown in the left panel of Fig.~\ref{fig-1} as a function of centrality, while the right panel displays the correlation between net-charge and net-kaon, $C_\mathrm{Q,K}$. Both correlations deviate from their Poisson baseline values, which are zero and one for $C_\mathrm{p,K}$ and $C_\mathrm{Q,K}$, respectively. The effects of resonance decays on these correlations are explored by comparing experimental measurements with Thermal-FIST \cite{TheFIST} model calculations that include and exclude resonance decay contributions. The model calculations use the canonical ensemble (CE) formalism, which enforces exact conservation of B, Q, and S in a correlation volume of $V_c$ = $3\mathrm{d}V/\mathrm{d}y$, with model parameters taken from Ref.~\cite{TheFISTparam}. The correlations $C_\mathrm{p,K}$ and $C_\mathrm{Q,K}$ are notably enhanced across all centralities due to resonance decays, and the model results incorporating these decay contributions align more closely with the data.

\begin{figure*}[h]
\centering
\includegraphics[width=0.48\linewidth,clip]{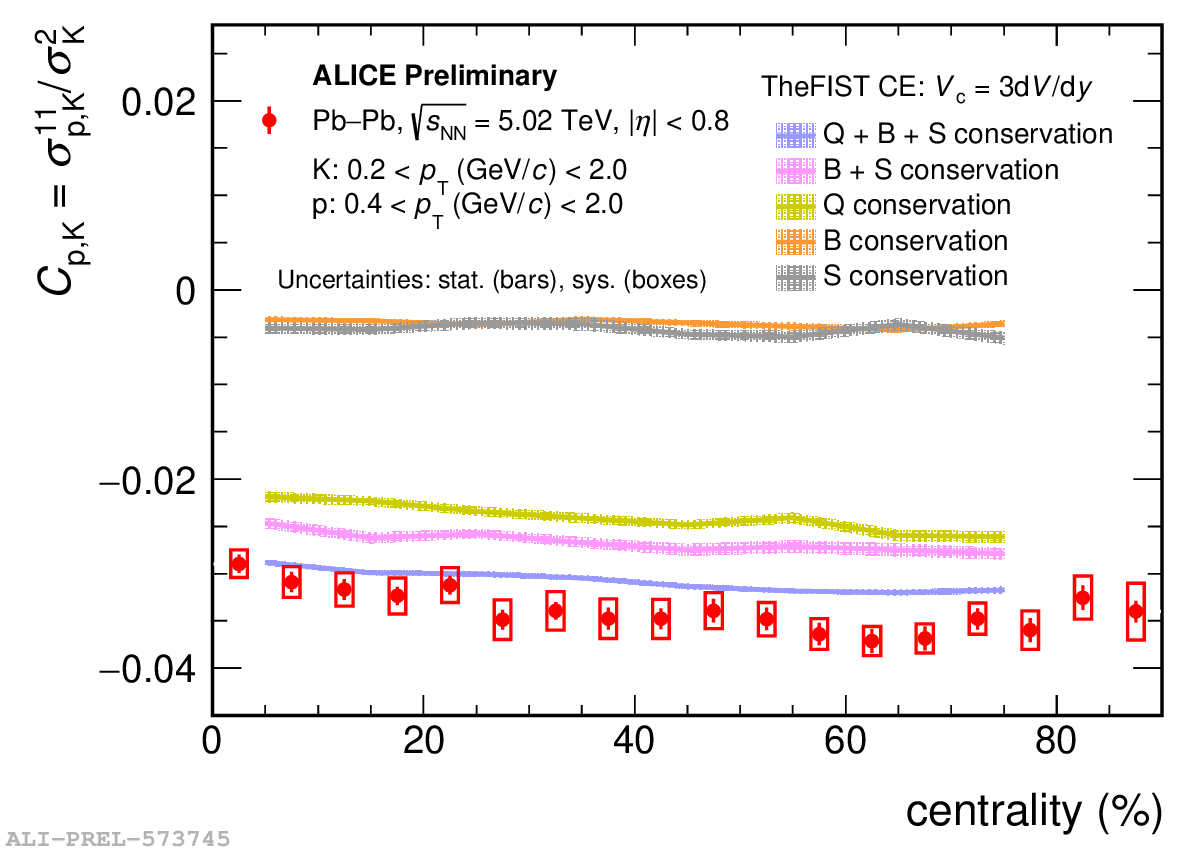}
\includegraphics[width=0.48\linewidth,clip]{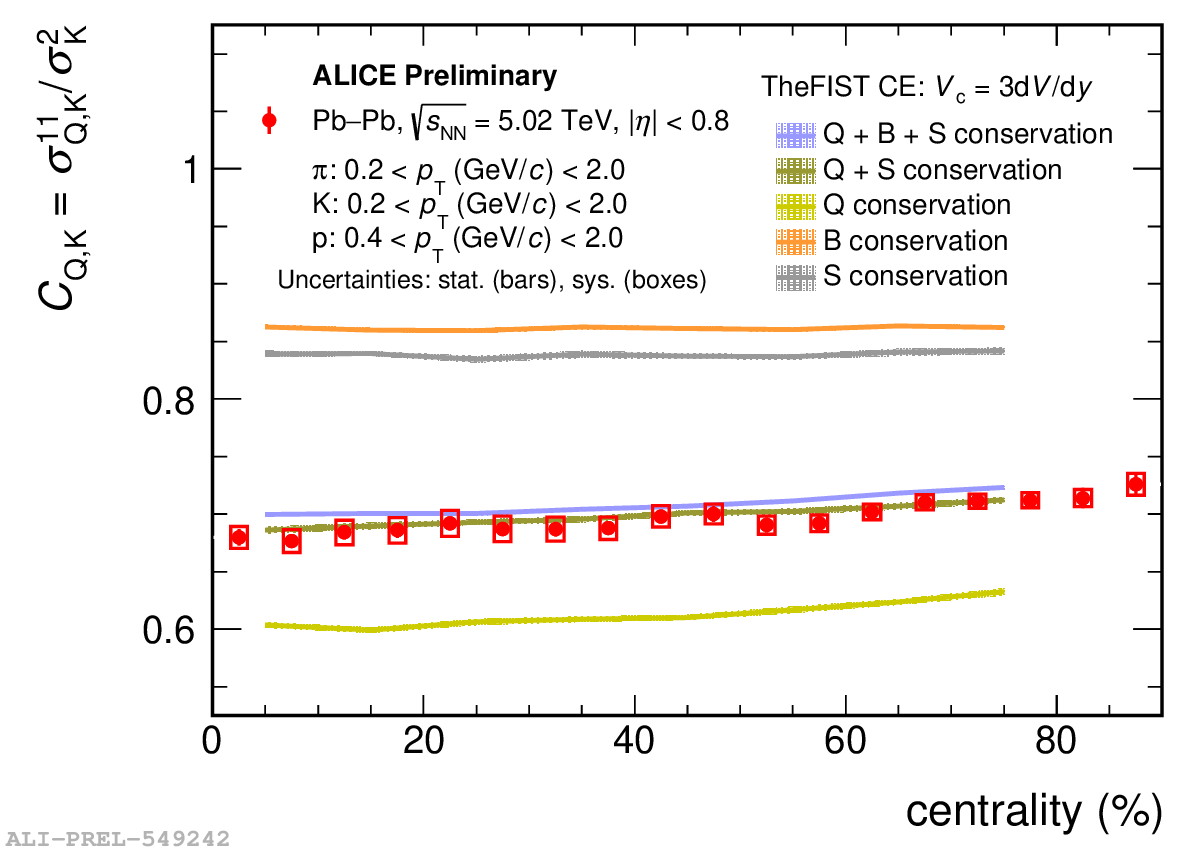}
\caption{Centrality dependence of $C_\mathrm{p,K}$ (left) and $C_\mathrm{Q,K}$ (right) in Pb--Pb collisions at $\sqrt{s_\mathrm{NN}} = 5.02$ TeV. ALICE measurements are compared with predictions from the Thermal-FIST (TheFIST) \cite{TheFIST} model in the canonical ensemble (CE), accounting for different conservation effects of electric charge (Q), baryon number (B), and strangeness (S).}
\label{fig-2}       
\end{figure*}

The conservation laws for Q, B, and S significantly influence the observed correlations in high-energy nuclear collisions. These laws constrain the types of particles produced, leading to distinct patterns in final-state particle correlations. The effect of different charge-conservation scenarios, investigated in the framework of Thermal-FIST model is reported here. In the left (right) panel of Fig.\ref{fig-2}, the experimental results for $C_\mathrm{p,K}$ ($C_\mathrm{Q,K}$) are compared with model predictions imposing the conservation of electric-charge-only, strangeness-only, baryon-number-only, both strangeness and baryon number (electric charge), and all three combined. $C_\mathrm{Q,K}$ is found to be mainly sensitive to conservation of both electric charge and strangeness. For $C_\mathrm{p,K}$, conservation of electric-charge-only, and both strangeness and baryon number have a similar effect. The mesaurements for both correlations are best described by model calculations, that account for the conservation of all three quantum numbers.

\begin{figure*}[h]
\centering
\includegraphics[width=0.48\linewidth,clip]{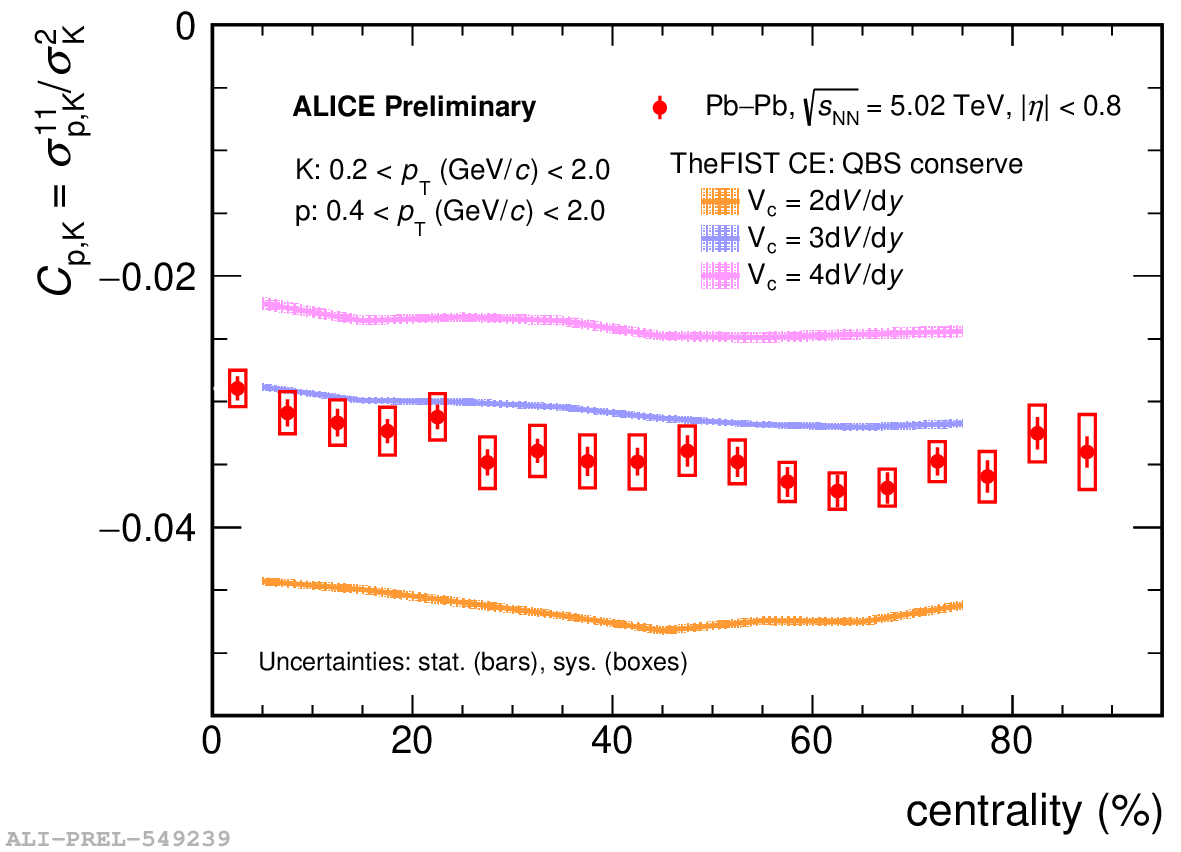}
\includegraphics[width=0.48\linewidth,clip]{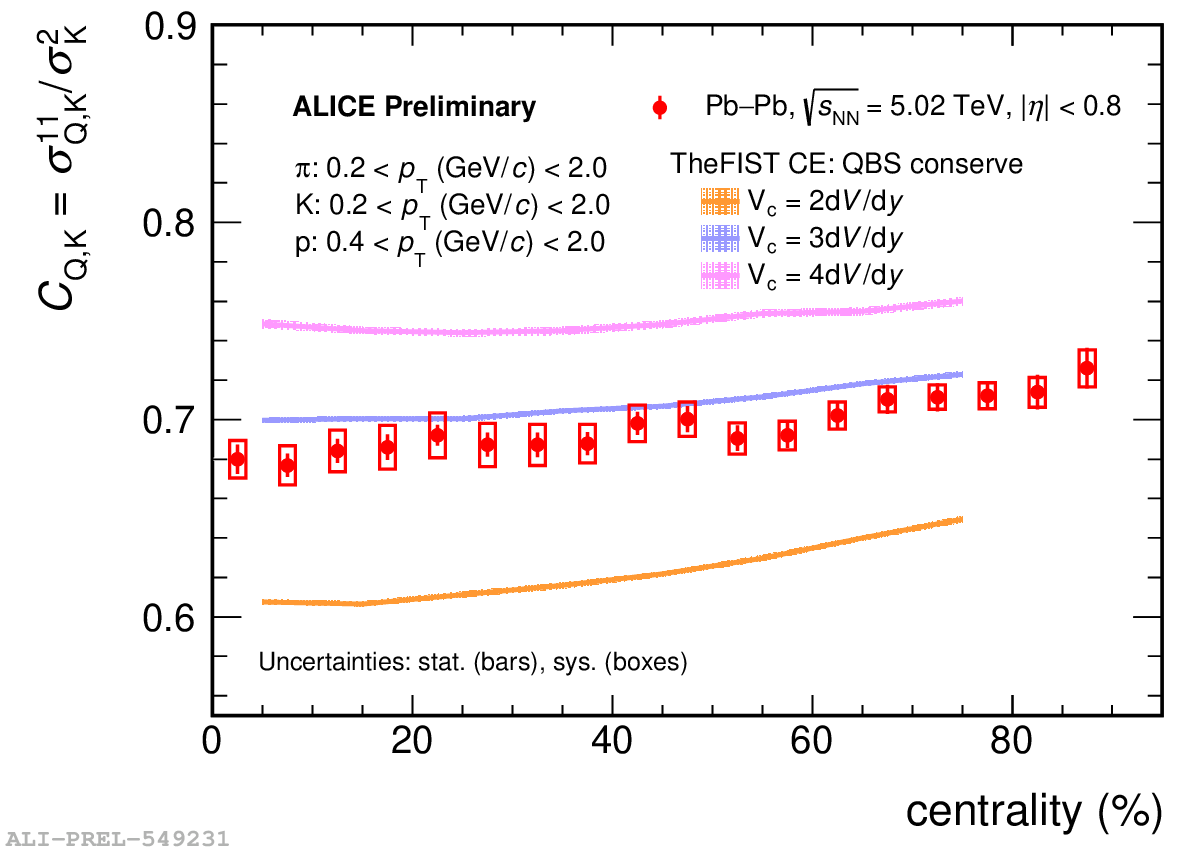}
\caption{Centrality dependence of $C_\mathrm{p,K}$ (left) and $C_\mathrm{Q,K}$ (right) in Pb--Pb collisions at $\sqrt{s_\mathrm{NN}} = 5.02$ TeV. ALICE measurements are compared with predictions from the Thermal-FIST (TheFIST) \cite{TheFIST} model in the canonical ensemble (CE), accounting for different correlation volumes, $V_c$ for conservation of electric charge (Q), baryon number (B), and strangeness (S).}
\label{fig-Vc}       
\end{figure*}

Figure \ref{fig-Vc} presents a comparison of experimental measurements for $C_\mathrm{p,K}$ and $C_\mathrm{Q,K}$ with Thermal-FIST model predictions for Q, B, S conservation across different correlation volumes, $V_c$. In the Thermal-FIST model, $V_c$ represents the finite spatial region within which particle correlations are established and charge conservation is imposed. The size and geometry of this volume significantly impact the observed correlation patterns. As $V_c$ decreases, deviations from the Poisson baselines (0 for $C_\mathrm{p,K}$ and 1 for $C_\mathrm{Q,K}$) become more pronounced. The agreement between model predictions and experimental data improves when $V_c$ is set to $3\mathrm{d}V/\mathrm{d}y$ (i.e., the volume of the system is approximately three times the rapidity density), suggesting that this volume adequately reflects the spatial extent and conservation constraints relevant to particle correlations observed in heavy-ion collisions. 

\section{Conclusions}
\label{conclusion}
In conclusion, the measured correlations between net-proton and net-kaon ($C_\mathrm{p,K}$) and between net-charge and net-kaon ($C_\mathrm{Q,K}$) in Pb--Pb collisions at $\sqrt{s_\mathrm{NN}} = 5.02$ TeV deviate significantly from their respective Poisson baseline values across all centralities. Resonance decays play a major role in enhancing these correlations, as demonstrated by the close agreement between experimental data and model calculations incorporating decay contributions. The Thermal-FIST model, which enforces exact conservation of baryon number, electric charge, and strangeness, also incorporates a correlation volume $V_c$--representing the spatial region over which these conservation laws are applied. The model results highlight the significant influence of both conservation laws and $V_c$ on the observed correlations. Overall, the measurements are better described when the model incorporates all three quantum number conservation laws within a correlation volume of $3\mathrm{d}V/\mathrm{d}y$.

\end{document}